\documentclass[aps,floatfix,footinbib, twocolumn,superscriptaddress,amsmath,amssymb,amsfonts]{revtex4-1}

\usepackage[T1]{fontenc}
\usepackage{lmodern} 
\usepackage{color}
\usepackage{bbold}
\usepackage{dsfont}
\usepackage{graphicx}
\usepackage[caption=false]{subfig}
\usepackage[colorlinks]{hyperref}
\usepackage[bottom]{footmisc}
\usepackage{enumerate}
\usepackage{float}

\begin{document}

\title{Optomechanical feedback cooling of a 5 mm-long torsional mode}

\author{Dianqiang Su}
\affiliation{State Key Laboratory of Quantum Optics and Quantum Optics Devices, Institute of Laser Spectroscopy, Shanxi University, Taiyuan 030006, People's Republic of China.}
\affiliation{Collaborative Innovation Center of Extreme Optics, Shanxi University, Taiyuan 030006,
People's Republic of China.}

\author{Yuan Jiang}
\affiliation{State Key Laboratory of Quantum Optics and Quantum Optics Devices, Institute of Laser Spectroscopy, Shanxi University, Taiyuan 030006, People's Republic of China.}
\affiliation{Collaborative Innovation Center of Extreme Optics, Shanxi University, Taiyuan 030006,
People's Republic of China.}

\author{Pablo Solano}
\affiliation{Departamento de F\'isica, Facultad de Ciencias F\'isicas y Matem\'aticas, Universidad de Concepci\'on, Concepci\'on, Chile.}
\affiliation{CIFAR Azrieli Global Scholars program, CIFAR, Toronto, Canada.}

\author{Luis A. Orozco}
\affiliation{Joint Quantum Institute, Department of Physics and NIST, University of Maryland, College Park, MD 20742, USA.}

\author{John Lawall}
\affiliation{National Institute of Standards and Technology,Gaithersburg MD 20899, USA.}

\author{Yanting Zhao}
\affiliation{State Key Laboratory of Quantum Optics and Quantum Optics Devices, Institute of Laser Spectroscopy, Shanxi University, Taiyuan 030006, People's Republic of China.}
\affiliation{Collaborative Innovation Center of Extreme Optics, Shanxi University, Taiyuan 030006,
People's Republic of China.}
\email{zhaoyt@sxu.edu.cn}

\begin{abstract}
We report three orders of magnitude optical cooling of the fundamental torsional mode of a 5 mm-long, 550 nm diameter optical nanofiber. The rotation of the nanofiber couples to the polarization of  guided laser fields. We use a weak laser probe to monitor the rotation, and use feedback to modulate the polarization of an auxiliary drive laser providing torque. Our results present a tool for the optomechanical control of large-scale torsional resonators, with metrological applications and potential implications for studying macroscopic objects in quantum states.
\end{abstract}

\maketitle


\section{Introduction}
Optomechanics uses light to monitor and control the motion of micro and macro-scale 
objects \cite{aspermeyer14}. State-of-the-art optical cooling has reached the quantum ground state of translational motion in a number of platforms 
\cite{OConnell2010,Verhagen2010,Chan2011,Teufel2011,Palomaki2013,Purdy2013,Safavi-Naeini2013,Rossi2018,Delic2020}
, an essential step for fundamental tests of quantum mechanics on massive objects 
\cite{Marshall2003,Marletto2017,Bose2017}. 
In such a context, larger and more massive 
systems will enable us to test the limits of current theories 
\cite{Schmole2016,AlBalushi2018,Liu2021,Westphal2021}. Moreover, precise control and 
transduction of mechanical motion enables metrological applications \cite{Schliesser2009}. 
Controlling rotational degrees of freedom, however, remains challenging \cite{Shi2016}, in part because rotation does not couple naturally to an optical cavity.

In this work, we report purely optical feedback cooling \cite{Vitali2002,aspermeyer14,Kleckner2006,corbitt07b,Gieseler2012,Dania2021} of a 5-mm long torsional resonator with a frequency of 190 kHz, reducing the mean-square angular displacement over three orders of magnitude from room temperature. The platform is the fundamental torsional mode of an optical nanofiber (ONF), coupled to the polarization of the guided light \cite{wuttke13,fenton18,Su2022}.
We perform in-loop and out-of-loop measurements,  and observe cooling from the reduction of the angular fluctuations and the broadening of the spectral density of the fluctuations. The measured optimal cooling is near the theoretical limit of the technique given by the signal-to-noise ratio (SNR) \cite{Poggio2007,Penny2021}, scaling as $\sim 2\sqrt{1/\rm{SNR}}\approx 1.2\times 10^{-3}$. Moreover, the platform presents a torque sensitivity $\sim 10^{-26}$ NmHz$^{-1/2}$, comparable to the record sensitivities achieved with nanodumbells \cite{Ahn2020}. Our results demonstrate ONFs to be a fruitful platform for rotational optomechanics, with potential applications in metrology and quamtum optomechanics.

 \begin{figure}
\begin{center}
\includegraphics[width=0.4\textwidth]{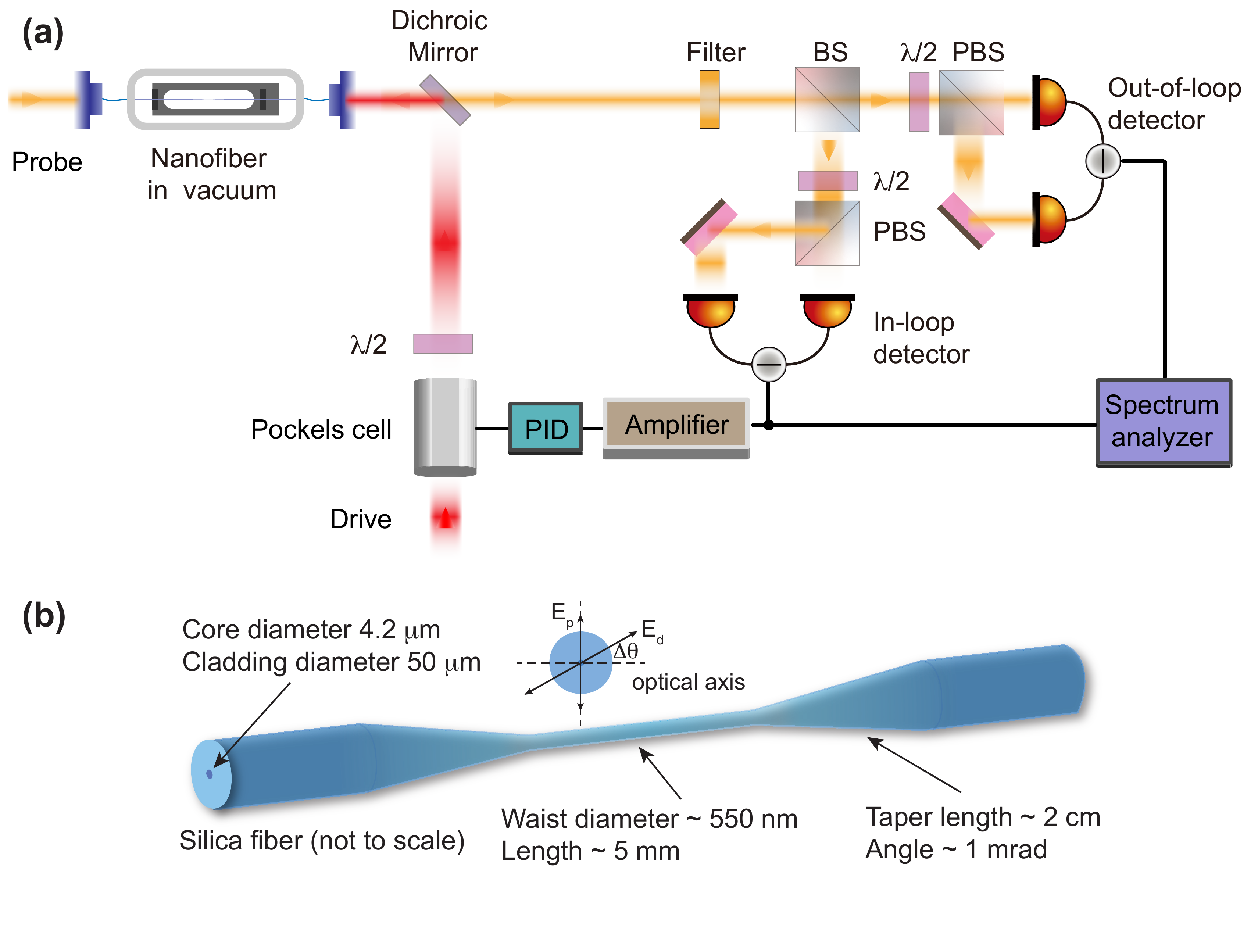}
\caption{{\bf(a)} Schematic of the apparatus. Probe and drive laser beams counter-propagate with independent polarization control. A beam splitter (BS) separates the probe in two for out-of-loop and in-loop detection. Each path has a half-waveplate ($\lambda$/2) to set the detection basis, followed by a polarizing beam splitter (PBS) and a balanced photodiode pair. The out-of-loop detection signal goes to a spectrum analyzer while the in-loop signal splits parts, one to the spectrum analyzer, and the other is amplified, filtered, and then goes to a control unit that slightly rotates the drive polarization closing the feedback loop. {\bf(b)} Schematic of the ONF with two effective polarization axis associated with ordinary and extraordinary indices of refraction, aligned with the optical axis, but at an angle $\Delta\theta$ with the input light polarization. The fiber is clamped (not shown) in the unmodified section.} 
\label{fb_apparatus}
\end{center}
\end{figure}

\section{Theoretical model}
The ONF is a silica cylinder of diameter 550~nm and length 5 mm created by tapering a length of standard optical fiber, as shown in Fig. \ref{fb_apparatus}. It has string, compressional, and torsional modes \cite{Hummer2019}, the latter of which couple to the polarization of guided light due to intrinsic birefringence produced during the fabrication process.  When linearly polarized light of power $P_{\text{opt}}$ propagates through the fiber, it results in an optically-induced torque~\cite{Su2022,friese98}
\begin{equation}
\tau_{\rm{opt}}=\tau_0 P_{\text{opt}}\sin(2(\theta-\theta_{\rm{L}}))
\label{eqn: torque}
\end{equation}
where~$\tau_0$ has units of torque per unit power, $\theta$ is the angle of the slow axis, and~$\theta_{\rm{L}}$ is the angle of the polarization.

We take the ONF to be driven by a fluctating Langevin torque $\tau_{th}$ with white spectral density, and an external torque from applied optical feedback.  
The equation of motion for the angular coordinate~$\theta$ representing the slow axis of the fundamental torsional mode is:
\begin{equation}
I\ddot{\theta}+\gamma \dot{\theta}+\kappa \theta=\tau_{\rm{th}}+\tau_0 P_{\text{opt}}\sin(2(\theta-\theta_{\rm{L}})),
\label{eqmotion}
\end{equation}
where $I$ is the effective moment of inertia of the mode, $\gamma$ is the damping coefficient, and $\kappa$ is the torsional spring constant.  $\tau_{\rm{th}}$ is the thermally induced torque with a double sided power spectral density $S_{\tau_{th}}=2\gamma k_B T$, where $T$ is the temperature and $k_B$ is the Boltzmann constant.  In the absence of optically-applied torque ($P_{\text{opt}}=0$), the system comes to thermal equilibrium with $\langle\theta^2 \rangle=k_BT/\kappa\sim 10^{-8}$ rad$^2$.

When a static optical field is introduced, the torque results in a new equilibrium angle $\bar{\theta}$, found by dropping the time derivatives and fluctuating torque~$\tau_{\rm{th}}$ in~(\ref{eqmotion}), yielding the transcendental equation $\kappa \bar{\theta}=\tau_0 P_{\text{opt}}\sin(2(\bar{\theta}-\theta_{\rm{L}}))$. Feedback will be introduced by modulating the polarization angle~$\theta_{\rm{L}}$ with a Pockels cell.  Taking $\theta_{\rm{L}}=\bar{\theta}_{\rm{L}}+\delta\theta_{\rm{L}}$ and $\theta=\bar{\theta}+\delta\theta$ and linearizing~(\ref{eqmotion}) about the steady-state, one obtains
\begin{equation}
I\delta\ddot{\theta}+\gamma\delta \dot{\theta}+\kappa \delta\theta=\tau_{\rm{th}}+2\beta\tau_0 P_{\text{opt}}
(\delta\theta-\delta\theta_{\rm{L}})
\label{eqfluctuations}
\end{equation}
where we have defined $\beta=\cos(2(\bar{\theta}-\bar{\theta}_{\rm{L}}))$. Taking the Fourier transform, we find
\begin{widetext}
\begin{equation}
\left(-\omega^2+i\omega\Gamma+\omega_m^2 -\frac{2\beta P_{\text{opt}}}{I}\tau_0[\omega]
\right)\delta\theta[\omega]=\frac{1}{I}\left(\tau_{\rm{th}}[\omega]-2\beta P_{\text{opt}}\tau_0[\omega]
\delta\theta_{\rm{L}}[\omega]\right),
\label{FTeqfeedbackmotion}
\end{equation}
\end{widetext}
where $\Gamma=\gamma/I$ and $\omega_m^2=\kappa/I$, and we have allowed for frequency dependence in~$\tau_0[\omega]$.  
In previous work \cite{Su2022} we demonstrated that the instrinsic delay in the response of the torque to changes in~$\theta_{\rm{L}}$, arising from the finite speed of sound, led to self-cooling with fixed optical drive ($\delta\theta_{\rm{L}}=0$).

Here, we use active feedback such that $\delta\theta_{\rm{L}}[\omega]=G[\omega]\delta\theta[\omega]$, where $G[\omega]$ describes the collective transfer function of the balanced photodetector, amplifier, PID controller, and Pockels cell shown in Fig.~\ref{fb_apparatus}.  In practice, there will always be measurement noise~$\theta_{n}$ to which the feedback will respond as well; taking $\delta\theta[\omega]\rightarrow \delta\theta[\omega]+\theta_{n}[\omega]$, we find
\begin{widetext}
\begin{equation}
\left(-\omega^2+i\omega\Gamma+\omega_m^2 -\frac{2\beta}{I}P_{\text{opt}}\tau_0[\omega](1+G[\omega])
\right)\delta\theta[\omega]=\frac{1}{I}\left(\tau_{\rm{th}}[\omega]-2\beta P_{\text{opt}}\tau_0[\omega]
G[\omega]\theta_n[\omega]\right),
\label{FTeqfeedbackmotion2}
\end{equation}
\end{widetext}

To proceed, we neglect the frequency dependence of~$\tau_0$ and take $G[\omega]=i\omega G_D$, corresponding to pure derivative feedback.  Eq.~(\ref{FTeqfeedbackmotion2}) then takes the form of a harmonic oscillator with optically-modified damping rate $\Gamma^{\prime}=\Gamma-\frac{2\beta}{I}P_{\text{opt}}\tau_0G_D$
and frequency $\omega_m^{\prime\,2}=\omega_m^{2}-\frac{2\beta}{I}P_{\text{opt}}\tau_0$, 
driven by fluctuations with torque spectral density
\begin{equation}
S_{\tau}=2I\Gamma k_B T+4\beta^2P_{\text{opt}}^2\tau_0^2G_D^2\omega^2 S_{\theta_n}[\omega]
\end{equation}
where $S_{\theta_n}$ is the spectral density of the measurement noise.

The spectral density of the angular fluctuations is then given by
\begin{equation}
S_{\delta\theta}=\frac{1}{I^2}\frac{2I\Gamma k_B T+4S_{\theta_n}[\omega]\beta^2P_{\text{opt}}^2\tau_0^2G_D^2\omega^2}{\left(\omega_m^{\prime\,2}-\omega^2\right)^2+\omega^2\Gamma^{\prime\,2}}
\label{Sfluct}
\end{equation}
Integration over all frequencies yields the mean-square angular fluctuations.  Assuming a white spectral density for the measurement noise, the integral can be evaluated analytically:
\begin{equation}
\langle \delta\theta^2\rangle=\frac{1}{2\pi}\int_{-\infty}^{\infty}S_{\delta\theta}\,d\omega=\frac{k_BT}{I}\frac{\Gamma}{\Gamma'}\frac{1}{\omega_m^{\prime\,2}}
+\frac{2\beta^2P_{\text{opt}}^2\tau_0^2G_D^2}{I^2\Gamma'}S_{\theta_n}
\end{equation}

Defining an effective mode temperature by $k_BT_{\text{mode}}=I\omega_m^{\prime\,2}\langle \delta\theta^2\rangle$ and a dimensionless feedback gain $g=-\frac{2\beta}{I\Gamma}P_{\text{opt}}\tau_0G_D$, one finds 
\begin{equation}
\frac{T_{\text{mode}}}{T}=\frac{1}{1+g}\left(1 +g^2\frac{S_{\theta_n}}{S_s}\right)
\label{Tmode}
\end{equation}
where $S_s=\frac{2k_BT}{\Gamma I\omega_m^{\prime\,2}}$
is the on-resonant spectral density of the angular fluctuations~(\ref{Sfluct}) in the absence of feedback. The dimensionless gain~$g$ can be varied by means of the polarization angle~$\bar{\theta}_{\rm{L}}$ (via~$\beta$), the optical power, or the electronic gain. By differentiating~(\ref{Tmode}), one finds that for a given SNR  $S_s/S_{\theta_n}$, the mode temperature is minimized for $g_{\text{opt}}=\sqrt{1+S_s/S_{\theta_n}}-1$.
In the limit of large SNR, $g_{\text{opt}}\rightarrow\sqrt{\frac{S_s}{S_{\theta_n}}}$ and
\begin{equation}
\frac{T_{\text{mode}}}{T}\rightarrow \frac{2}{\sqrt{S_s/S_{\theta_n}}}
\label{theorylimit}
\end{equation}
This shows that the SNR is the crucial figure of merit for cooling.

The reduction of the mode temperature can be measured in various ways.  Most fundamentally, it is defined by $k_BT_{\text{mode}}=I\omega_m^{\prime\,2}\langle \delta\theta^2\rangle$, where $\langle \delta\theta^2\rangle$ is determined in terms of the integral of the measured distribution~$S_{\delta\theta}$.  It is also related to the broadened linewidth~$\Gamma^{\prime}$ of $S_{\delta\theta}$ by $T_{\text{mode}}/T=\Gamma/\Gamma^{\prime}\left(1+\frac{1}{SNR}(\Gamma^{\prime}/\Gamma-1)^2\right)$, so that the cooling scales inversely with the linewidth of the angular spectral density as long as the linewidth broadening is not too great; for large enough values of the feedback, the linewidth will continue to broaden but the mode temperature will rise.  Finally, the squared fluctuating amplitude~$\delta\theta^2(t)$ of the torsional oscillation can be measured in the time domain, and the statistics of a long series of measurements will follow a Boltzmann distribution
\begin{equation}
   p(\delta\theta^2)=\frac{\kappa}{2 k_B T_{\rm{mode}}}e^{-\kappa\,\delta\theta^2/(2 k_B T_{\rm{mode}}) }
\label{MB} 
\end{equation}
from which $T_{\text{mode}}$ can be extracted.

\section{Experimental setup}
Figure~\ref{fb_apparatus} shows the experimental apparatus. We heat and pull \cite{hoffman14} a commercial optical fiber (Fibercore SM1500) to produce a 550 nm diameter, 5 mm length waist with a 1 mrad taper. For the wavelengths used, it allows the propagation of the fundamental HE$_{11}$ mode \cite{ravets13}. The nanofiber resides in a vacuum chamber evacuated with an ion pump to suppress air damping. The fundamental torsional resonance of the ONF is at about 190 kHz with a half width at half maximum (HWHM) of $0.75(5)$ Hz, corresponding to a mechanical quality factor $Q\approx 1.26(8) \times 10^5$. 

In order to measure the angular fluctuations of the ONF, $250 \mu$W of linearly polarized probe laser light at 852 nm is coupled in, and the transmitted polarization is redundantly analyzed by two pairs of balanced photodetectors (BPD).  Rotation of the ONF causes a linear rotation of the output polarization due to the ONF birefringence, and the balanced detection scheme provides a signal proportional to $\delta\theta$ while removing common-mode laser intensity fluctuations.  One BPD is used as an ``in-loop’’ detector for feedback, and the other is used for ``out-of-loop`` detection.   The signal from either BPD can be sent to a spectrum analyzer, used either to observe the spectral density of the signal, or in zero-span mode as a fixed programmable bandpass filter to eliminate technical noise in a measurement of the squared angular fluctuations $\delta\theta^2(t)$ as a function of time.  Calibration is accomplished by assuming that the observed fluctuations in the absence of feedback are of thermal origin at room temperature.

Feedback is applied by means of a linearly polarized ``drive'' laser (Fig.~\ref{fb_apparatus}(a)) whose polarization angle~$\theta_L$ is controlled by a Pockels cell, generating a torque on the ONF as given by  ~(\ref{eqn: torque}). The output of one of the ``in-loop’’ BPD is amplified and filtered so as to produce a signal corresponding approximately to derivative feedback, and then applied to the Pockels cell.  The loop gain can be controlled by means of the mean polarization~$\bar{\theta}_{\rm{L}}$, the drive laser power, or the electronic gain.  We observe similar results in each case, but vary the drive laser power in this work.

\begin{figure}[t]
\begin{center}
\includegraphics[width=0.4
\textwidth]{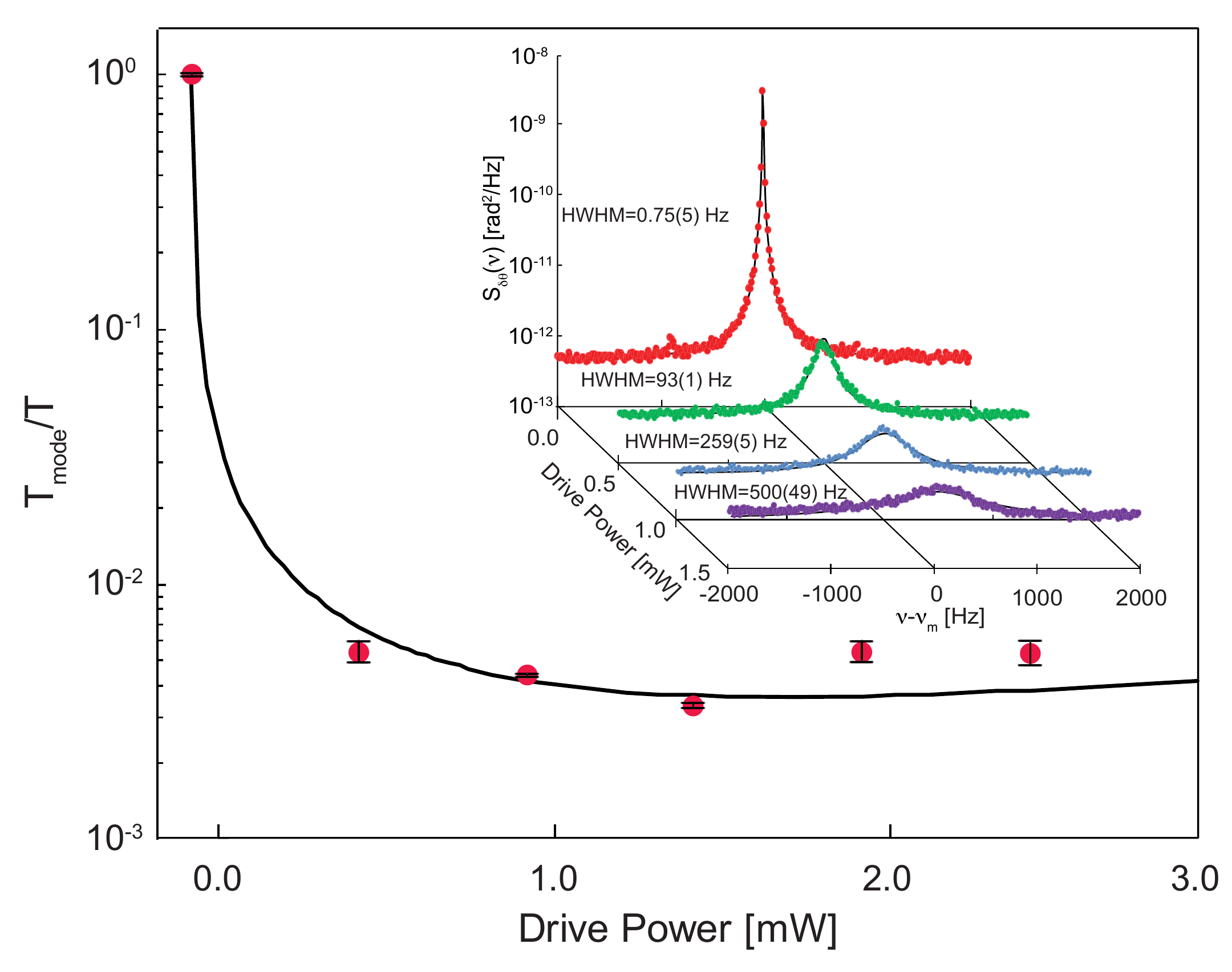}
\caption{Mode temperature, calculated from the integral of~$S_{\delta\theta}(\omega)$ as determined from out-of-loop measurements, as a function of the drive laser power.  The continuous line is a fit to Eq.~(\ref{Tmode}). The inset shows the evolution of~$S_{\delta\theta}(\omega)$ as the drive power is raised.} 
\label{fig:PSD}
\end{center}
\end{figure}

\section{Results}
Figure \ref{fig:PSD} shows the ratio of the mode temperature to the ambient temperature as a function of drive power (feedback).  The mode temperature is inferred from the integral of the out-of-loop angular spectral density~$S_{\delta\theta}(\omega)$, and the solid line shows a fit to Eq.~(\ref{Tmode}).  The inset shows the evolution of ~$S_{\delta\theta}(\omega)$ as the drive power increases. The amplitude drops towards the noise floor, the width broadens, and the mean-square angular fluctuation $\langle \delta\theta^2\rangle$ given by the integral of~$S_{\delta\theta}$ diminishes.  Drive powers higher than 1.5~mW do not result in better cooling.  Indeed, ``squashing'' \cite{Poggio2007} (not shown) appears in the in-loop signal as the~$g^2$ term in Eq.~\ref{Tmode} takes over, rendering the cooling less effective. The lowest mode temperature that we infer from these data is $T_{\text{mode}}/T=4.0(1)\times10^{-3}$.

\begin{figure}[t]
\begin{center}
\includegraphics[width=0.4
\textwidth]{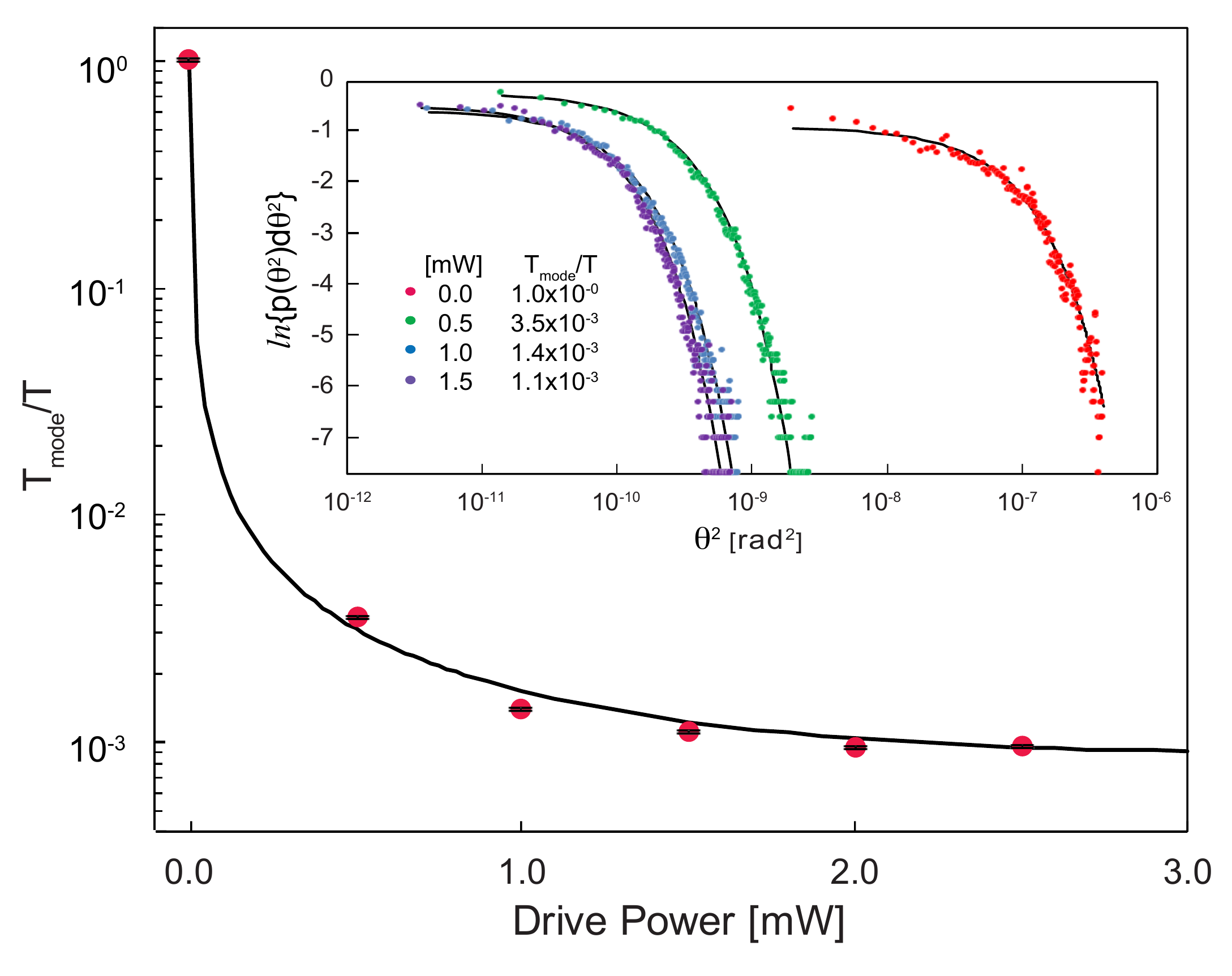}
\caption{Mode temperature calculated from the measured statistical distribution of the mean-square angular fluctuation $\langle \delta\theta^2\rangle$ as a function of drive laser power.  The continuous line is a fit to Eq.~(\ref{Tmode}). The inset shows representative measured distributions $p(\theta^2)$ and their fits to Eq.~(\ref{MB}).  The fits would appear linear on semilogarithmic axes, but making the horizontal axis logarithmic as well facilitates the visualization of the reduction of~$\langle \delta\theta^2\rangle$.
} 
\label{fb_MBIL}
\end{center}
\end{figure}

Figure~\ref{fb_MBIL} shows a complementary measurement of the mode temperature, made by using the spectrum analyzer as a combination of square-law detector and bandpass filter to infer the squared fluctuating amplitude~$\delta\theta^2(t)$. Again the measurements use out-of-loop data, and the solid line is a fit to Eq.~\ref{Tmode}.  Representative
statistics of $\delta\theta^2$ for time series measurements of 10~s are shown in the inset, along with fits to the Boltzmann distribution given in Eq.~(\ref{MB}).   
The limiting temperature that we observe here is $T_{\text{mode}}/T=1.11(1) \times10^{-3}$. The uncertainties correspond to one standard deviation as obtained using the $\chi^2$ method of the fits.

\section{Discussion and Outlook}
It is of interest to compare the degree of cooling that we achieve to the expected limit given by Eq.~(\ref{theorylimit}) from the signal to noise ratio. The noise floor is dominated by shot noise and electronic (dark) noise, both of similar amplitudes for a 250 $\mu$W probe. We measure the contribution of both noise sources to the system by measuring the electronic voltage noise without the ONF, both with and without 250 $\mu$W of probe light striking the detectors.  We refer the voltage noise back to effective angular noise by dividing it by the same calibration constant used to interpret the data with the ONF.  The amplitude of the signal for the data set shown in Fig. \ref{fig:PSD} measured on resonance is $S_s=4.50(9)\times10^{-9}$ rad$^2$/Hz, while for the data set shown in  Fig. \ref{fb_MBIL} is $S_s=\frac{2\langle\delta\theta^2\rangle}{\Gamma}=3.90(3)\times10^{-8}$ rad$^2$/Hz. The difference is due to systematic experimental variations typically observed. The nanofiber system is made from a non-polarization-maintaining optical fiber, which causes the light polarization at the ONF to drift, and imposes a technical challenge to set truly linearly polarized light at the ONF waist \cite{Joos2019}.  The differences in the signals from the data sets shown here highlight the role of the SNR in the cooling performance. The corresponding limiting mode temperature, from Eq. (\ref{theorylimit}), is $T_{\text{mode}}/T=4.90(4) \times 10^{-3}$ and $T_{\text{mode}}/T=1.664(4) \times 10^{-3}$ respectively, in close agreement with the observed values. To increase the signal to noise ratio, it would be desirable to enhance the mechanical transduction.  Since the birefringence supplying the transduction in the ONF was an unintended artifact of the fabrication process, increasing it by a modification of the process seems plausible.

An independent study, performed in parallel to ours, shows similar results using electrodes for feedback-cooling an ONF torsional mode \cite{Tebbenjohanns2023}.

Beyond the optomechanical cooling capabilities of the platform, its high sensitivity to rotations makes ONFs a viable candidate for a torque sensor once systematic effects are controlled. The tensile strength of an ONF could allow its use as a torsional pendulum for precision force measurement \cite{Westphal2021}. Torsional modes also couple to external electric fields \cite{wuttke13,Tebbenjohanns2023} presenting a potential field sensor. The sensitivity of the platform is ultimately defined by the noise floor, $S_{\theta_n}$, corresponding to a rotational sensitivity of $\approx 1.6\times10^{-7} ~\rm{rad}/\sqrt{Hz}$. The conversion from rotation to torque depends on the modulus of the angular displacement susceptibility, which on resonance is $\chi(\omega_0)=1/\omega_0\gamma$. We thus obtain a torque sensitivity of $\approx 2.9 \times 10^{-26}$ NmHz$^{-1/2}$, competitive with state of the art rotational sensors \cite{Ahn2020}.  The large scale of the system could allow for a larger interaction region of the sensor, improving the overall sensitivity and enabling measurements of quantum vacuum friction of polarizable objects near surfaces \cite{Ahn2020}.

The ground state energy of the torsional harmonic oscillator corresponds to a temperature of $\approx9~\mu$K. Since ONFs are compatible with dilution refrigerators in the mK range \cite{Voigt2015}, it seems plausible to enter the quantum regime with a combination of cryogenic and optical feedback cooling. Thus ONFs appear to offer promise as a candidate to study quantum torsional optomechanics of relatively massive ($\approx ng$) and large-scale ($\approx$~cm size) objects.

\section{Conclusion}
In sumamry, we demonstrate optical feedback cooling of the fundamental torsional mode of a 5-mm long optical nanofiber, reducing the effective mode temperature by three orders of magnitude, reaching a mode temperature of $\approx 320$ mK using optomechancial transduction in a cavityless system. The polarization of the guided light couples to the fiber via its intrinsic birefringence, enabling both a sensitive probe of its rotation and a mechanism to optically apply torque for control purposes.  
We characterize the cooling in both the frequency and time domains, and find results that are near the limit imposed by the signal to noise ratio. Finally, we discuss the possibilities of utilizing the platform as a torque sensor, and suggest that reaching the quantum regime should be possible with a combination of cryogenic and optical feedback cooling.

{\it Acknowledgment.---} We thank Peter Zoller for his continuous interest.
P.S. is a CIFAR Azrieli
Global Scholar in the Quantum Information Science Program. This work was supported by the National Key Research and Development Program of China (No. 2022YFA1404201), National Natural Science Foundation of China (Nos. 12034012, 12274272, 61827824, 62105191, 12074231), Fundamental Research Program of Shanxi Province(20210302124537), ``1331 KSC'', PCSIRT (No. IRT\_17R70), 111 Project (No. D18001),  CONICYT-PAI 77190033, and FONDECYT 11200192 from Chile.

{\it Disclosures.---} The authors declare no conflicts of interest.

{\it Data Availability Statement.---} Data underlying the results presented in this paper are not publicly available at this time but may be obtained from the authors upon reasonable request.

\bibliography{Torsion}

\end{document}